\documentclass[aip,preprint]{revtex4}
\usepackage{graphicx}
\usepackage{dcolumn}
\usepackage{bm}
\usepackage[english]{babel}

\begin{document}
\title{Nonlocal supercurrent in mesoscopic multiterminal SNS Josephson junction in the low-temperature limit}
\author{T.~E.~Golikova$^{a}$\email{golt2@list.ru},
M.~J.~Wolf$^{b}$, D.~Beckmann$^{b}$, I.~E.~Batov$^{a}$, I.~V.~Bobkova$^{a}$, A.~M.~Bobkov$^{a}$, and V.~V.~Ryazanov$^{a,c}$}
\affiliation{$^{a}$Institute of Solid State Physics RAS, 142432
Chernogolovka, Moscow district, Russia\\$^{b}$Institute of Nanotechnology, Karlsruhe
Institute of Technology, 76021 Karlsruhe, Germany\\$^c$Russian Quantum Center, Novaya str.,100, BC ``URAL'', SKOLKOVO, Moscow region, 143025, Russia}
\begin{abstract}
A nonlocal supercurrent was observed in mesoscopic planar SNS Josephson junctions with additional normal-metal electrodes, where nonequilibrium quasiparticles were injected from a normal metal electrode into one of the superconducting banks of  the Josephson junction in the absence of a net transport current through the junction. We claim that the  observed effect is due to a supercurrent counterflow, appearing to compensate for the quasiparticle flow in the SNS weak link. We have measured the responses of SNS junctions for different distances between the quasiparticle injector and the SNS junction at temperatures far below the superconducting transition temperature. The charge-imbalance relaxation length was estimated by using a modified Kadin, Smith, and Skocpol scheme in the case of a planar geometry. The model developed allows us to describe the interplay of charge imbalance and Josephson effects in the nanoscale proximity system in detail.      
\end{abstract}

\maketitle
\section{Introduction}
\indent Nonlocal effects in superconducting mesoscopic systems have attracted a lot of interest in the recent decade in connection with investigations of novel coherent effects such as crossed Andreev reflection (CAR) and elastic cotunneling (EC) \cite{deutscher2000,BeckmannCAR, Beckmann2, russo2005, Chandrasekhar}, spin diffusion, injection, and accumulation \cite{Jedema1, Jedema2, johnson1994,Haviland,poli2008} that could have future practical applications in superconducting spintronics. Modern superconducting electronic nanodevices often operate under out-of-equilibrium conditions because their sizes are comparable with relaxation lengths. Complex biased circuits can contain Josephson junctions in the resistive state, as well as normal-metal or ferromagnetic elements that are sources of nonequilibrium quasiparticles. One of the first nonlocal nonequilibrium effects was observed in a quite macroscopic Josephson SNS sandwich-type junction many years ago \cite{RyazanovPL, Kap_Ryaz_Schmidt}. The SNS structure used had a thin S electrode with a thickness comparable with the charge imbalance length, $\lambda_{Q^{*}}$, and was excited to the nonequilibrium state by quasiparticle injection. Charge imbalance in superconductors was investigated in detail and attracted a lot of interest both in experimental and theoretical studies in the 1970s and 1980s. However, it was investigated mostly at temperatures close to the superconducting transition temperature $T_{c}$, both experimentally \cite{Clarke, Yu_Mercereau, RyazanovJLTP, vanHarlingen, Nad, Santh} and theoretically \cite{Tinkham_Clarke,Schmid_S,Artemenko_Volkov}. The low-temperature range has been studied experimentally only recently \cite{yagi2006,Hubler,Chandrasekhar,kleine2010,arutyunov2011} and still has no appropriate theoretical description. While experiments with tunnel injection report on relaxation lengths of a few $\mu$m, in agreement with older results close to $T_{c}$, results for Ohmic injection \cite{Chandrasekhar} poorly correlate with 
previous results at high temperatures, where $\lambda_{Q^{*}}$ is determined only by the inelastic electron-phonon time \cite{vanHarlingen, Nad}. To measure the charge-imbalance length, authors of Refs. \onlinecite{Chandrasekhar} and \onlinecite{Hubler}  used normal-metal probes contacted to superconducting films directly or via a tunnel junction following the first works \cite{Clarke, Yu_Mercereau, vanHarlingen} and recent spin injection experiments \cite{Jedema1, Jedema2}. Another nonlocal detection method, proposed in \cite{RyazanovPL, Kap_Ryaz_Schmidt}, uses a Josephson junction as a sensor of the injected nonequilibrium quasiparticles. 
This first ``nonlocal Josephson effect'' was observed in sandwich-type SNS structures also at temperatures very close to $T_{c}$. An advantage of this method, which we also use in this work, is given by the possibility to measure directly the quasiparticle flow. Moreover, this phenomenon furnishes the clue to a nonlocal control of Josephson junctions in modern superconducting electronic nanodevices.            

\begin{figure}[tbp]
\centering
\includegraphics[width=0.47\textwidth]{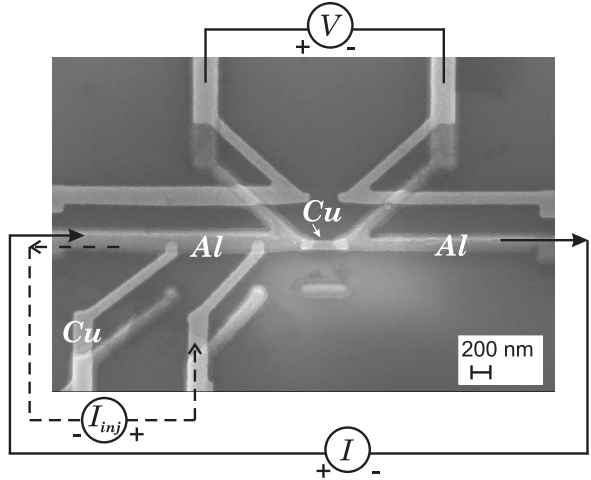}
\caption{SEM image of an Al-Cu-Al multiterminal Josephson junction with two Cu-strips (injectors) connected to the left junction bank together with
the local (solid line) and nonlocal (dashed line)  measurement schemes.}
\label{scheme}
\end{figure}

In this paper, we present the detection of a ``nonlocal critical current'' in mesoscopic SNS (Al-Cu-Al) Josephson junctions with several spatially separated normal metal injectors (Cu) connected to one of the superconducting Al banks (Fig.~\ref{scheme}). To measure the nonlocal voltage at low temperatures $T\ll T_c$, we use superconducting leads just near the junction. 
Besides, in order to describe the interplay of charge imbalance and Josephson effect in the realized mesoscopic system 
in the low temperature range, we have elaborated a two-channel charge imbalance model proposed previously in \cite{Kap_Ryaz_Schmidt, KSS}.     

\section{Experiment}
Figure \ref{scheme} shows a scanning electron microscopy (SEM) image of one of our
samples, together with a scheme of the measurement setup. The submicron-scale multiterminal planar structures
were fabricated by means of electron beam lithography and \textit{in situ} shadow evaporation. First, a copper layer with a thickness of $d_{N}$=30 nm was deposited onto an oxidized silicon substrate to create the weak link of the Josephson Al-Cu-Al junction as well as the two Cu-injectors. Without breaking the vacuum, a thick  aluminum layer with a thickness of around  100 nm was evaporated at a second angle in order to form all superconducting leads, 
so that all SN interfaces are assumed to be highly transparent. The injectors were realized as sloped narrow strips of a width of around 100 nm in order to avoid a Josephson coupling between the left superconducting bank of the SNS junction and the aluminum shadow parallel to the injectors (see Fig.~\ref{scheme}). The investigated structures had nominally identical SNS Josephson junctions, characterized by the same distance \textit{L}=250 nm between superconducting electrodes, but had different distances \textit{d} between the injectors and  the SNS junction. Type-I
samples (Fig. \ref{scheme}) had two injectors with \textit{d} equal to approximately 1.5 and 0.5 $\mu$m, while for type-II samples, these distances were 1.5 and 1.0 $\mu$m. Such a choice of the distance scale allows us to exclude coherent CAR and EC effects \cite{BeckmannCAR} as well as the inverse proximity effect, because the distances were much larger than the Al coherence length $\xi_{S}\approx$130 nm \cite{myJETP}. On the other hand, nonequilibrium phenomena, characterized by the charge-imbalance length ($\lambda_{Q^{*}}\sim1 \mu$m \cite{Chandrasekhar}), remain to be investigated with this choice of distances $d$. 

All transport
measurements in local and nonlocal configuration were performed using a standard four-terminal method in a shielded
cryostat at temperatures down to 0.37 K. Two stages of resistor-capacitor (RC) filters
were incorporated into the measurement dc lines to eliminate electrical
noise: a warm RC-filter stage was located at the top of the cryostat, while a cold RC-filter stage was mounted directly at the sample holder. 

\begin{figure}[tbp]
\centering
\includegraphics[width=0.47\textwidth]{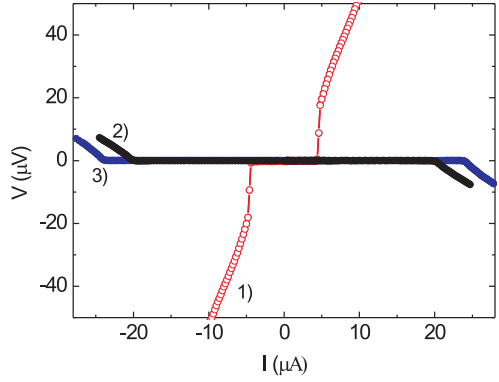}
\caption{(Color online) Current-voltage characteristics of an Al-Cu-Al Josephson junction (sample A1) at $T$=0.4 K. Curve 1, conventional local measurement; curve 2, nonlocal mearurement, current is injected from the nearest ($R$) Cu strip; curve 3, nonlocal measurement, current is injected from the farther ($L$) Cu strip.}
\label{iv}
\end{figure}

Figure \ref{iv} represents the current-voltage characteristics at \textit{T}=0.4 K for one of our samples measured in local and nonlocal configurations shown in Fig. \ref{scheme}. 
In the local configuration, the transport current is fed to the Josephson SNS junction directly through its horizontal superconducting aluminum  leads. The voltage across the Josephson junction is measured via two additional superconducting leads outside the current path in a four-terminal configuration. In this case, we observe the conventional IV curve which is typical for such submicron SNS Josephson junctions \cite{Courtois, myJETP} with the critical current $I_{c} \simeq$ 4 $\mu$A (Fig. \ref{iv}). At currents slightly exceeding the critical current $I_{c}$, a sharp voltage increase is observed, but the characteristic is still not hysteretic. In the nonlocal configuration, the current is passed from one of the normal metal injectors to the left side of the horizontal aluminum  lead bypassing the Josephson junction (Fig. \ref{scheme}), while the voltage is measured across the Josephson junction, as in the case of the local configuration. Curves 2) and 3) represent the 
nonlocal current-voltage characteristics for 
injection from the nearest, ``right'' ($R$), and the farther, ``left'' ($L$), injectors with \textit{d}=0.5 and 1.5 $\mu$m, correspondingly. In the nonlocal case, registered critical currents were larger than for the local configuration: $I_{cR}^{inj}=19.7 \mu$A and $I_{cL}^{inj}=23.6 \mu$A (for sample A1) and the voltage had opposite sign. A detailed model of the nonlocal effect and its discussion will be presented in the subsequent section. We give only a short description here.  A significant part of the quasiparticles injected from normal metal to the superconductor has an energy larger than the superconducting gap $\Delta$ and hence penetrates into the superconductor. The charge imbalance and appropriate decrease of the pair chemical potential $\mu_s$ result in a longitudinal electric field $E= \nabla \mu_s$/$e$ (where $e$ is the charge of the electron). The charge-imbalance length $\lambda_{Q^{*}}$ is a characteristic length of the conversion of the nonequilibrium quasiparticles to pairs, i. e., the penetration length of quasiparticle flow in both directions from the normal injector.  The quasiparticles penetrate also through the Josephson junction into the right superconducting bank if the distance \textit{d} between the injector and the Josephson junction is less than $\lambda_{Q^{*}}$. The total current through the Josephson junction is zero, so a counterflow of Cooper pairs has to arise. It compensates for the quasiparticle flow until it reaches the critical current. Besides,
the nonlocal voltage $V$, equal to a drop of the pair electrochemical potential, has opposite sign as compared with the local experiment for the current directions shown in Fig. \ref{scheme}. The observed ``injection critical current'' $I_{c}^{inj}$ is much larger than $I_{c}$, since only a small part of the current from the injector reaches the Josephson junction due to Andreev reflection at the NS-interface and the quasiparticle conversion between the injector and the Josephson junction. For the same reason, the injection critical current $I_{cL}^{inj}$ from the farther injector is larger than the $I_{cR}^{inj}$ value from the nearest one.

\section{Model of a Josephson SNS junction in the case of quasiparticle injection}

Nonequilibrium processes in a superconductor including the conversion of a quasiparticle flow into a pair current can be described reasonably by means of an equivalent circuit introduced by Kadin, Smith, and Skocpol (KSS) \cite{KSS} for the explanation of phase-slip-center behavior. The KSS approach was simplified for the case of low-frequency processes and extended to study the effect of  nonequilibrium quasiparticle flow on Josephson SNS junctions by Kaplunenko, Ryazanov, and Schmidt \cite{Kap_Ryaz_Schmidt}. A similar model modified for the geometry of our planar Josephson structures and low-temperature range is developed in this section. 

The longitudinal electric field $E$ originating from the quasiparticle charge imbalance $Q^{*}$ arises in nonequilibrium regions of a superconductor where the quasiparticle current $I_n$ converts into the current of Cooper pairs $I_s$:
\begin{equation}
div \mathbf{I_n} = -div \mathbf{I_s} \propto U,\enspace \nabla U = - E.
\label{div_eq}
\end{equation}

The gauge-invariant potential $U$ is related to the deviation of the pair chemical potential $\delta\mu_s$ from its equilibrium value ($U = -\delta \mu_s/e \propto Q^{*}$) and can be detected directly by normal and superconducting probes placed in the nonequilibrium region \cite{Yu_Mercereau, Dolan_J}. Relaxation of $Q^{*}, U$, and $I_n$ is characterized by the charge-imbalance length $\lambda_{Q^{*}}$ \cite{Artemenko_Volkov}:

\begin{equation}
\lambda_{Q^{*}}^{2}\nabla^{2}Q^{*}= Q^{*},\enspace \lambda_{Q^{*}}^{2}\nabla^{2}U=U,\enspace \lambda_{Q^{*}}^{2}\nabla^{2}I_n= I_n.
\label{nabla_eq}
\end{equation}

\begin{figure}[tbp]
\centering
\includegraphics[width=0.47\textwidth]{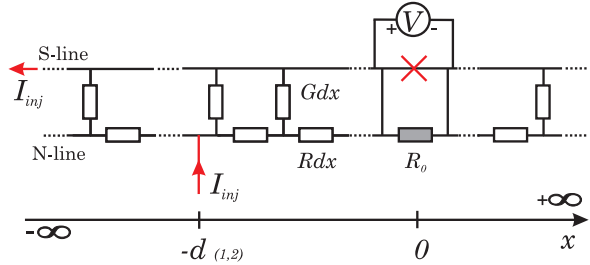}
\caption{(Color online) Scheme of the equivalent circuit used in calculations. Central part is a Josephson SNS junction (the cross), $R_{0}$ is the resistance of Cu-weak link, \textit{d} is a distance between N injector and Josephson junction.}
\label{eq}
\end{figure}

For the discussed experiment (Fig.\ref{scheme}), a one-dimensional model can be used. An equivalent circuit shown in Fig.~\ref{eq} represents a two-fluid approach taking into account two distinct electrochemical potentials for the normal (quasiparticle) and superconducting components. We choose the junction center as the origin of coordinates in our one-dimensional problem and the direction of the \textit{x} axis parallel to the horizontal Al leads of the junction. The resistive line (N-line) represents a channel for the quasiparticle current $I_{n}$ and the potential of this line corresponds to the electrochemical potential of the quasiparticles. Here, \textit{R} is the normal-state resistance of the aluminum lead per unit length and $R_{0}$ is the resistance of the Cu bridge of the Josephson junction. In turn, the superconducting line (S line) is a channel for the supercurrent $I_{s}$, and the potential of this line is the electrochemical potential of the Cooper pairs. The cross in the scheme marks the 
region of the 
junction phase-slipping and the pair electrochemical potential drop is measured by the voltmeter connected by two superconducting leads. The ``conductances'' $G$ describe channels of the quasiparticle conversion into the condensate.
Kirchhoff's circuit laws give the following equations for the equivalent circuit element (Fig.~\ref{eq}):

\begin{equation}
\frac{dI_n}{dx}+GU=0,\enspace \frac{dU}{dx}+RI_n=0,\enspace \frac{d^2I_n}{dx^2}= GRI_n. 
\label{Kirch_eq}
\end{equation}
  
A comparison of Eqs.~\ref{nabla_eq} and Eqs.~\ref{Kirch_eq} shows that the conversion of the quasiparticles into the condensate per unit length is determined by the ``conductance'' $G=(R\lambda_{Q^{*}}^{2})^{-1}$. We assume the lengths of the horizontal aluminum  arms to be infinite in our model. De facto in our experiment, the lengths of the left and right arms were approximately 8 $\mu$m, i.e., essentially larger than $\lambda_{Q^{*}}$ and all distances $d$ between injectors and Josephson junction. Because of Andreev reflection at the SN interface, the current $I_{inj}$ passing through the normal injector is divided into normal and superconducting parts upon entering the superconductor. We introduce a coefficient $\beta=I_{n,inj}/I_{inj}$ to take into account the fraction of injected quasiparticles that do not convert to pairs due to Andreev reflection at the injector/superconductor interface. 

The equation for the normal current $I_n$ [see Eqs. (\ref{nabla_eq}) and (\ref{Kirch_eq})]
\begin{equation}
\frac{d^2I_n}{dx^2}-\lambda_{Q^{*}}^{-2} I_n=0
\label{in_eq}
\end{equation}
should be solved under the appropriate boundary conditions at the injector point ($x=-d$) and at the Josephson junction $x=0$ of the equivalent circuit shown in Fig.~\ref{eq}:
\begin{eqnarray}
I_n(x=-0)=I_n(x=+0)  \nonumber \\
I_n (x=-d+0)+I_s(x=-d+0)=0  \nonumber \\
I_s(x=-d-0)+I_n(x=-d-0)=-I_{inj} \nonumber \\
I_s(x=-d-0)=I_s(x=-d+0)-(1-\beta) I_{inj} \nonumber \\
U(x=-d-0)=U(x=-d+0) \nonumber \\
U(x=+0)-U(x=-0)-V+I_n(x=0)R_0=0
\label{bc}
\enspace . 
\end{eqnarray}

Here, the first equation is just the continuity of the current flowing via the N line at the Josephson junction, the second and the third equations represent the conservation of the total current, and the fourth equation is the continuity of the current at the injector ($x=-d$). $U(x)=-\frac{1}{G}\frac{dI_n}{dx}$ [see Eq.~(\ref{Kirch_eq})] is the difference between the chemical potentials of the normal and superconducting components, that is, the potential difference between the N line and the S line. In the last equation, $V$ is the electrochemical pair potential difference at the Josephson junction. Solving Eq.~(\ref{in_eq}) supplemented by boundary conditions (\ref{bc}), one can obtain the following Josephson equation       

\begin{equation}
\frac{\hbar \dot{\varphi}}{2e\widetilde{R}}+I_{c}sin\varphi=-\frac{R\lambda_{Q^{*}}\beta I_{inj}e^{-d/\lambda_{Q^{*}}}}{\widetilde{R}}
\label{JJ}
\enspace ,
\end{equation}

where $\widetilde{R}$=$R_{0}$+2$R\lambda_{Q^{*}}$, $V \equiv \frac{\hbar \dot{\varphi}}{2e}$ and $\varphi$ is the phase difference across the Josephson junction. From Eq.~(\ref{JJ}) we get the value of the injection critical current $I_{c}^{inj}$:
\begin{equation}
\frac{I_{c}}{I_{c}^{inj}}=\frac{R\beta \lambda_{Q^{*}}e^{-d/\lambda_{Q^{*}}}}{\widetilde{R}}
\label{lQ}
\end{equation}

This equation has a clear physical meaning. The right part is the fraction of the current from the injector which reaches the SNS junction. This fraction is proportional to $\beta$, an exponential factor, and the ratio $R\lambda_{Q^{*}}$/${\widetilde{R}}$. The coefficient $\beta$ is a kind of quasiparticle transmission coefficient at the injector/superconductor interface, since  $(1-\beta)I^{inj}$ is the part of the current from the injector that converts to pairs due to Andreev reflection at the NS interface. The ratio $R\lambda_{Q^{*}}/{\widetilde{R}}$ plays the role of a geometrical factor and determines the fraction of quasiparticles which go to the right side, because this right arm has the larger resistance $R_{0}$+$R\lambda_{Q^{*}}$ as compared with the left side resistance $R\lambda_{Q^{*}}$. This fraction undergoes an exponential decay between the injector and the SNS junction due to charge-imbalance relaxation. In the discussed  experiment, the main quantitative contribution to Eq.~(\ref{lQ}) is given by the 
geometrical factor.
If the coefficient $\beta$ is known for the particular conditions at the interface between the normal injector and the superconductor, one can, in principle, estimate the value of $\lambda_{Q^{*}}$ from Eq.~(\ref{lQ}). However, in order to calculate $\beta$, one should know a number of parameters which are not known exactly from the experiment (see discussion below). Therefore, we use another method to obtain $\lambda_{Q^{*}}$ from the experimental data. Assuming that the NS interface parameters corresponding to the left and the right injectors are approximately the same, one can find from Eq.~(\ref{lQ}):
\begin{equation}
\lambda_{Q^{*}}=\frac{d_L-d_R}{\log (I_{cL}^{inj}/I_{cR}^{inj})}
\label{lQ_ratio}
\enspace .
\end{equation}
We believe that this method gives quite reliable values of $\lambda_{Q^{*}}$ because it does not contain an uncertainty connected to the factor $\beta$ (Andreev reflection at the injector/superconductor interface) as well as to the unknown factor related with Andreev reflection processes for quasiparticles at the Josephson SNS junction \cite{Kap_Ryaz_Schmidt}. The only assumption, which is really used in Eq.~(\ref{lQ_ratio}), is the exponential decay of the normal quasiparticle flow in the superconductor. Strictly speaking, the equivalent scheme described above is only valid close to the critical temperature. Nevertheless, it is obtained theoretically by us \cite{bobkov}, that at low temperatures, the spatial behavior of the charge imbalance relaxation can be approximately considered as exponential. An exponential decay of charge imbalance at low temperatures was also found experimentally \cite{Hubler}. Therefore, it is quite reliable to extract $\lambda_{Q^{*}}$ from the experimental data according to Eq.~(\ref{lQ_ratio}) even at low temperatures.

\section{Results and discussion}

\begin{table} [b]
\caption{Characteristic parameters of three samples (A-C) and the results of estimations of $\lambda_{Q^{*}}$ and $\beta$ at $T$=0.4 K. Subindex L corresponds to the farther injector and R corresponds to the nearest one.}
\label{t}
\begin{tabular} {| c | c | c | c | c | c | c | c | c | c |}
\hline      
& $d_{L}$ & $d_{R}$& \textit{R} & $R_{0}$ & $I_{c}$ & $I_{cL}^{inj}$ & $I_{cR}^{inj}$ & $\lambda_{Q^{*}}$ & $\beta$ \\
& ($\mu$m) & ($\mu$m)& ($\Omega$/$\mu$m) & ($\Omega$) & ($\mu$A) & ($\mu$A)& ($\mu$A)& ($\mu$m) &\\
\hline A & 1.47& 0.56&0.65& 5.5 & 4.4 & 24 &20& 4.99 $\pm$ 0.31 & 0.91\\
\hline B & 1.53&0.59&0.69&5.6&4.4&23.6&19.7&5.20 $\pm$ 0.31 & 0.89\\
\hline C & 1.48&1.02&1.07&6.0&3.7&24.2&22.4&5.95 $\pm$ 0.71& 0.58\\
\hline
\end{tabular}
\end{table}

In Table \ref{t}, results for $\lambda_{Q^{*}}$ obtained by means of Eq.~(\ref{lQ_ratio}) are presented along with the necessary parameters extracted from the experiment. All values were measured at a temperature of 0.4 K. The estimated values of  $\lambda_{Q^{*}}\simeq 5\mu$m are in reasonable agreement with the values for aluminum obtained before at T$\ll$T$_c$  \cite{Hubler}.
Furthermore, the quasiparticle transmission coefficient $\beta$ at the injector/superconductor interface was estimated using the values of $\lambda_{Q^{*}}$ for each sample and Eq.~(\ref{lQ}). The magnitude of $\beta$ shows that the largest part of the current is injected in the superconductor as quasiparticle flow and the dominant process described by Eq.~(\ref{lQ}) is related to the conversion in the superconductor.   
For all measured samples, the charge-imbalance relaxation length was essentially larger than the distances $d_{L}$ and $d_{R}$ between injectors and Josephson junctions. Therefore, in accordance with our model, injected quasiparticles can not be converted entirely into pairs before the Josephson junction. As a result, the normal current and the longitudinal electric field penetrate into the region of the weak link and to the right-side bank of the junction where they eventually decay completely. The counterflow of Cooper pairs arises to compensate for the flow of quasiparticles, such that the total current on the right side of the injectors has to be zero. The critical injection current $I_{cL}^{inj}$ from the farther injector is higher than the value $I_{cR}^{inj}$ from the nearest one because the quasiparticles have to overcome a longer distance and create the same counterflow to reach the junction critical value of $I_{c}$. It is not very obvious for the moment that the proposed method of the $\lambda_{Q^{*}}$ measurement has 
clear advantages over the technique used in Refs. \cite{Hubler, Chandrasekhar} because it may well be that it also requires corrections related to a difference between coefficients $\beta_L$ and $\beta_R$ due to different voltages at the left and right N/S interfaces. Nevertheless, our experiment and model calculation give a description for real modern submicron Josephson circuits where nonequilibrium effects are undesirable or, on the contrary, can be used to tune Josephson junction characteristics.      

\begin{figure}[tbp]
\centering
\includegraphics[width=0.5\textwidth]{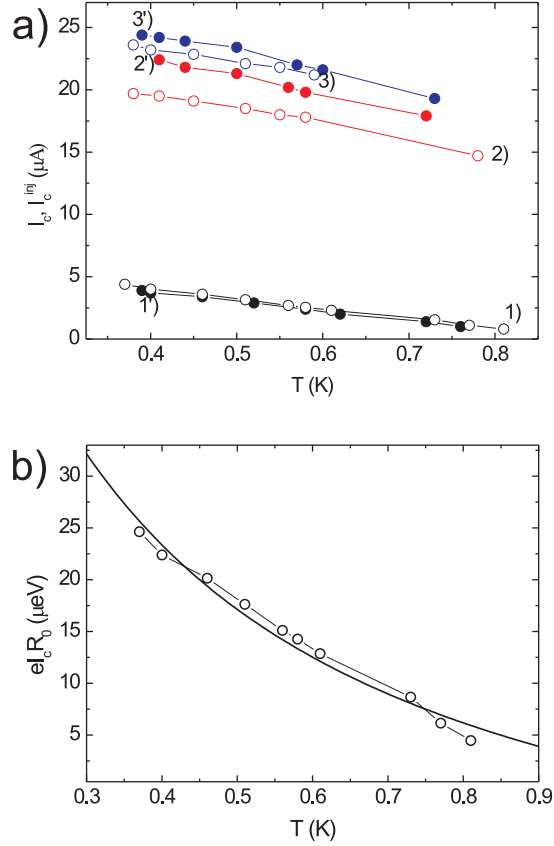}
\caption{(Color online) (a) Temperature dependences of the critical currents $I_{c}$ and the critical injection currents $I_{c}^{inj}$ for the samples B (open symbols) and C (closed symbols). Curves 1, 2, 3 are local, nonlocal $R$ (nearest injector), nonlocal $L$ (farther injector) cases for the sample B, correspondingly; and 1', 2', 3' are the same for the sample C. (b) Temperature dependence of the measured $eI_{c}R_{0}$ product for sample B (open symbols) and the corresponding theoretical fit (solid line).}
\label{IcT}
\end{figure}

To clarify the charge-imbalance mechanism and the nonequilibrium quasiparticle  distribution, we have studied the temperature dependencies of the critical currents and $\lambda_{Q^{*}}$. Temperature dependencies of the critical current $I_{c}$ and critical injection currents $I_{cL}^{inj}$ and $I_{cR}^{inj}$ for samples B and C are shown in Fig. \ref{IcT}(a). The temperature dependence of the conventional (local) critical current, $I_{c}$, shown in Fig. \ref{IcT}(b), is typical for this type of Al-Cu-Al junctions \cite{Courtois,myJETP}. The estimation of  the Thouless energy for our samples gives the value $E_{Th}=\hbar D_{n}/L^{2}$=80 $\mu$eV, where the copper-layer diffusion coefficient $D_{n}$ of about 80 cm$^{2}$/s was determined from resistive measurements in our previous work \cite{myJETP}. In comparison with the superconducting gap of the aluminum layer $\Delta$=180 $\mu$eV, $E_{Th}$ is smaller in our case, but not small enough to satisfy the long-junction limit \cite{Courtois,Dubos}. In order to 
make a theoretical fit of the product $eI_{c}R_{0}$ versus temperature for the local case, an approximation proposed for low temperatures in Ref. \onlinecite{Dubos} was used: $eI_{c}R_{0}/E_{Th}=a(1-be^{-aE_{Th}/3.2k_{B}T})$, where $a$=10.87 and $b=1.3$ were obtained in the long-junction limit ($\Delta\gg E_{Th}$) and for highly transparent SN interfaces. The best fit, presented in Fig. \ref{IcT}(b) by a solid line, gives $a$=0.87, $b$=1.25 for $E_{Th}$=80 $\mu$eV in our case. 

So $aE_{Th}$=$eI_{c}R_{0}(T=0)$=70 $\mu$eV instead of about 866 $\mu$eV [see Eq.~(2) in Ref.~\onlinecite{Dubos} for the long-junction limit]. Such a divergence is fully explicable. The use of the approximation by Dubos $et. al.$ is quite popular, not only in the long-junction limit and for transparent SN-interfaces. 
A similar discrepancy for the ratio $aE_{Th}$=$eI_{c}R_{0}(T=0)$ was discussed in several works \cite{Hammer, Carillo, Frielinghaus} and was associated with nonideal interfaces and the intermediate length of the junction $L\sim \sqrt{\hbar D_n/\Delta}$. A sharp change of the $eI_{c}R_{0}$ product with the increase of $E_{Th}/\Delta$ is presented in Fig.~1, Ref. \onlinecite{Dubos}. The interface barrier is taken into account by using a reduced effective Thouless energy as one more fitting parameter \cite{Hammer}. Our SNS junction parameters yield $E_{Th}/\Delta \simeq$ 0.4 and $r=R_b /R_N \simeq$ 0.2, where $R_b$ and $R_N$ are the interface and the copper strip resistance, correspondingly. (It should be noted that $R_N=R_0-2R_b$.) Estimates have shown that with these two factors one can obtain a decrease of $eI_{c}R_{0}$ by more than an order of magnitude.         
The critical current $I_{c}$ is identical for the two samples, while the farther the injectors are from the junction, the larger the nonlocal critical current $I_{c}^{inj}$ is in the whole investigated temperature range from 0.37 up to 0.8 K.

\begin{figure}[tbp]
\centering

\includegraphics[width=0.47\textwidth]{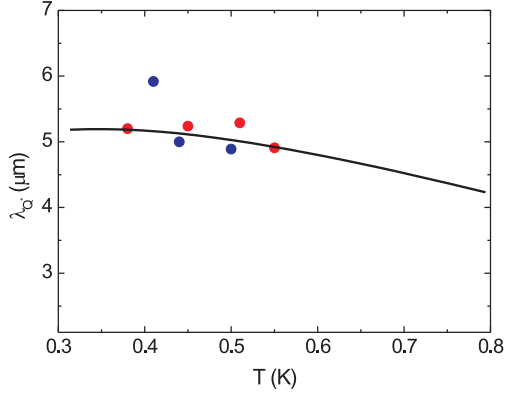}
\caption{Temperature dependence of $\lambda_{Q^{*}}$ for sample B data (red points) and sample C data (blue points) calculated using Eq.~(\ref{lQ_ratio}). The solid curve is the result of theoretical calculations.}
\label{lQ_T}
\end{figure} 

The temperature dependence of the charge-imbalance length, calculated from the experimental data using Eq.~(\ref{lQ_ratio}), is represented in Fig.~\ref{lQ_T}. In contrast to a sharp increase of $\lambda_{Q^{*}}(T)\propto(1-T/T_c)^{-1/4}$ observed close to $T_c$ \cite{Schmid_S, Artemenko_Volkov, Dolan_J, Nad, Kap_Ryaz_Schmidt}, the obtained temperature dependence is much weaker. Qualitatively, the observed weak decrease of $\lambda_{Q^{*}}$ with temperature increase can be explained as follows. In the diffusive case $\lambda_{Q^{*}} = \sqrt{D_s\tau_Q}$, where $D_s$ is the electron-diffusion constant and $\tau_Q$ is the charge-imbalance relaxation time. Near the critical temperature we have $\tau_Q \propto \tau_\varepsilon k_B T/\Delta(T)$, where $\tau_\varepsilon$ is the energy relaxation time and $\Delta(T)/k_B T$ represents a rough estimate for the fraction of quasiparticles which take part in the charge-imbalance relaxation. Taking into account that the energy relaxation time can be considered as 
temperature independent near $T_c$, it is seen that the temperature dependence of $\lambda_{Q^{*}}$ is only determined by the temperature dependence of $\Delta^{-1/2} \propto (1-T/T_c)^{-1/4}$. This result is well known \cite{Schmid_S,Clarke,Tinkham_Clarke}. 

In the intermediate temperature range, corresponding to our experimental data, the temperature dependence of $\Delta$ does not play the main part anymore. The fraction of the quasiparticles contributing to the charge-imbalance relaxation is of order unity. We assume that the charge-imbalance relaxation is only due to the inelastic electron-phonon scattering. Then, the main contribution to the temperature dependence of $\tau_Q$ is given by the temperature dependence of $\tau_\varepsilon$. It is well known, that $\tau_\varepsilon \propto T^{-3}$ at $\varepsilon < T$ \cite{agd}. Roughly speaking, this temperature behavior of  $\tau_\varepsilon$ is responsible for the observed weak decrease of $\lambda_{Q^{*}}$($T$) for the presented temperature range. However, to obtain the quantitative dependence, one should consider the temperature dependence of the energy gap $\Delta$($T$), the temperature and energy dependence of the inelastic electron-phonon time $\tau_\varepsilon$, and the energy dependence of the effective 
diffusion constant $D$$_s$ in a superconductor on an equal footing. In order to take into account all these factors properly, we have solved the kinetic equation derived in the framework of the quasiclassical Usadel equation \cite{usadel}. In contrast to the earlier consideration \cite{Schmid_S}, we do not restrict ourselves to temperatures close to $T_c$ in the particular calculation of $\lambda_{Q^{*}}$($T$) and take into account the explicit temperature dependence of $\tau_\varepsilon$. It allows us to explain the observed weak decrease of $\lambda_{Q^{*}}$($T$) taking place for intermediate temperatures. The detailed calculation will be published elsewhere \cite{bobkov}. The obtained theoretical dependence of $\lambda_{Q^{*}}$($T$) is plotted in Fig.~\ref{lQ_T} together with the experimental data. The only fitting parameter we use is the strength of the electron-phonon interaction. The particular value, which we take to fit the experimental data, corresponds to $\tau_\varepsilon(T=T_c)=3.6$~ns. This 
value agrees fairly well with the earlier 
experiments \cite{vanHarlingen,chi}.  It is seen that our theoretical results are in reasonable agreement with the experimental data. Nevertheless, we can not exclude contributions of other mechanisms to the charge-imbalance relaxation due to the elastic impurity scattering in the presence of gap anisotropy (nonuniformity) \cite{Tinkham} and supercurrent \cite{Galaiko}.           
     
\section{Conclusion}
To conclude, we have observed experimentally a nonlocal supercurrent in planar multiterminal SN structures appearing due to quasiparticle injection to one of the superconducting leads of a Josephson SNS junction. The observed effect  opens a new way to provide a nonlocal control of the Josephson current at the nanoscale. To describe the interplay of charge imbalance and the Josephson effect in the nanoscale Josephson system with normal-metal injectors, we have elaborated a two-fluid model appropriate for the nonequilibrium situation. In particular, the model was used to extract the charge-imbalance length $\lambda_{Q^{*}}$ and obtain its temperature dependence at low temperatures. Peculiarities of nonequilibrium phenomena in mesoscopic planar SN structures at low temperature are discussed. The experiment and model realized can be extended to more complicated nonequilibrium cases with superposition of charge and spin imbalances in SF nanostructures with superconductors and ferromagnets.
   
We acknowledge A.V. Ustinov  for support and useful discussions. The work was supported by grants of the Russian Academy of Sciencies and the Russian Foundation for Basic Research.

\end{document}